\documentclass{article}
\usepackage{epsfig}
\makeatletter
\def\eqalign#1{\null\,\vcenter{\openup\jot\m@th\ialign{\strut\hfil$%
        \displaystyle{##{}}$&$\displaystyle{{}##}$\hfil\crcr#1\crcr}}\,}
\makeatother
\begin{document}
\def\beq{\begin{equation}}
\def\eeq{\end{equation}}
\title{{\bf Fermi's "Golden Rule" and Non-Exponential Decay}}
\author{C. Dullemond\\Institute for Theoretical Physics,\\University of Nijmegen,\\Nijmegen, The Netherlands}
\maketitle

\begin{abstract}
A study is made of the behavior of unstable states in simple models which nevertheless are realistic representations of situations occurring in nature. It is demonstrated that a non-exponential decay pattern will ultimately dominate decay due to a lower limit to the energy. The survival rate approaches zero faster than the inverse square of the time when the time goes to infinity.
\end{abstract}
\section{Introduction}
In this article a study is made of the nature of the decay of unstable states in a nonrelativistic setting. One would expect this decay always to be exponential after some time, like the radioactive decay one finds in nature. 
It is the purpose of this article to demonstrate in a simple model with a ground state that next to a possible nondecaying bound state contribution there will be a non-exponential contribution to the decay pattern which will ultimately dominate the exponential decay. In practice this non-exponential contribution is extremely small and under normal conditions undetectable. 
Moreover it will arise only when one is able to keep a state coherent for a long time, which is not the case normally.
This phenomenon is not unknown and is discussed in general terms in the literature\footnote{J.J. Sakurai, "Modern Quantum Mechanics, Revised Edition", Supplement II (Addison Wesley, 1994); G-C. Cho, H. Kasari and Y. Yamaguchi, Prog. Theor. Phys. {\bf 90}, 803, (1993)}, but a model demonstration may be helpful for its understanding.

Starting point is a simplified model giving rise to exact exponential decay. This will be worked out in Section 2. In Section 3 the model is modified as to make it more realistic. 
This modified model nevertheless leads to a superposition of exponential decay contributions of which one will ultimately dominate. Both models have an essential flaw which is caused by the fact that no absolute lower limit to the energy has been imposed. If an energy lower limit is taken into account then a non-exponential term appears to be unavoidable. 
Section 4 gives a discussion of this effect.
\section{A simple, exactly solvable model}
Consider a free Hamiltonian \(H_{0} (=H_{0}^{\dagger})\), a set of kets \({\{}|E{\rangle}{\}}\) and a particular ket \(|a{\rangle}\) with the properties:
\begin{equation}H_{0}|E{\rangle}=E|E{\rangle},\hspace{1cm} H_{0}|a{\rangle}=\alpha|a{\rangle},\hspace{1cm}-\infty<E<\infty\end{equation}
\begin{equation}{\langle}E|E'{\rangle}=\frac{1}{\rho}\delta(E-E'),\hspace{1cm}{\langle}E|a{\rangle}={\langle}a|E{\rangle}=0,\hspace{1cm}{\langle}a|a{\rangle}=1\end{equation}
\begin{equation}\rho\int_{-\infty}^{\infty}|E{\rangle}{\langle}E|dE+|a{\rangle}{\langle}a|=I,\hspace{1cm}\rho>0\end{equation} 

Here \(\rho\) is the density of free energy-eigenstates, assumed to be independent of \(E\). Furthermore, there is an interaction Hamiltonian \(H'(=H'^{\dagger})\) with the properties:
\beq H'|E\rangle =a|a\rangle,\hspace{1cm}H'|a\rangle =\rho a \int_{-{\infty}}^{{\infty}} |E\rangle dE\eeq 

Here we take \(a\) to be real, positive and independent of \(E\). We have: 
\beq \langle a|H'|E\rangle =\langle E|H'|a\rangle=a \eeq 

This is the transition matrix element.

Let the total Hamiltonian \(H\) be defined as:
\beq
H=H_{0}+H'
\eeq
We are interested in expressions of the form:
\beq \langle a|e^{-\frac{it}{\hbar}H}|a\rangle \eeq

These are transition matrix elements for the transition of a state at time \(t=0\) to the same state at time \(t\geq 0\). Traditionally, one reasons as follows: Up to time \(t=0\) the total Hamiltonian is \(H_{0}\) which allows an experimenter to prepare a system in a pure energy eigenstate \(|a{\rangle}\). Then, at time \(t=0\) the "perturbation" \(H'\) is "switched on" and at a later time \(t\) "switched off". From that time on one can carry out an analysis of the resulting state, in particular one can try to find out what the chances are that the original state is found back. According to Heisenbergs uncertainty principle for time and energy it takes an infinite amount of time to prepare the original state and to analyse the final results, but the available time to do so is indeed unlimited. The method of time dependent perturbation theory leads then to the famous "Golden Rule" of Fermi.\footnote{See any textbook on quantum mechanics.}

In order to evaluate these matrix elements we put \(H\) in its spectral form: 
\beq H= \int_{-\infty}^{\infty}\lambda |\tilde{\lambda}\rangle \langle \tilde{\lambda}|d\lambda \eeq \beq H|\tilde{\lambda}\rangle =\lambda|\tilde{\lambda}\rangle \eeq \beq \langle \tilde{\lambda}|\tilde{\lambda}'\rangle=\delta(\lambda - \lambda') \eeq \beq \int_{-\infty}^{\infty}|\tilde{\lambda}\rangle \langle \tilde{\lambda}|d \lambda =I \eeq 

Here we have made the assumption that \(H\) has no discrete eigenkets. 

We make the following expansion: 
\beq |\tilde{\lambda}\rangle=\int_{-\infty}^{\infty} f_{\lambda}(E)|E\rangle dE+c_{\lambda}|a\rangle \eeq 

The normalization condition then gives:
\beq \frac{1}{\rho} \int_{-\infty}^{\infty}f_{\lambda}^{*}(E)f_{\lambda '}(E)dE+c_{\lambda}^{*}c_{\lambda '}=\delta (\lambda -\lambda ') \eeq 

Next we solve the equation: 
\beq (H_{0}+H'-\lambda )|\tilde{\lambda} \rangle =0 \eeq 

We have:
\beq
\eqalign{
&(H_{0}+H'-\lambda )|\tilde{\lambda} \rangle \cr
&=(H_{0}+H'-\lambda )[ \int_{-\infty}^{\infty}f_{\lambda}(E)|E\rangle dE+c_{\lambda} |a\rangle ]\cr
&=\int_{-\infty}^{\infty}f_{\lambda}(E)[(E-\lambda )|E\rangle +a|a\rangle ]dE\cr&\qquad+c_{\lambda} [(\alpha -\lambda )|a\rangle +\rho a \int_{-\infty}^{\infty}|E\rangle dE]=0 
}
\eeq 

This can be satisfied only if: \beq (E-\lambda )f_{\lambda}(E)+\rho a c_{\lambda}=0\eeq and \beq (\alpha -\lambda )c_{\lambda}+a\int_{-\infty}^{\infty}f_{\lambda}(E)dE=0 \eeq so that \beq f_{\lambda}(E)=- \rho a c_{\lambda}P\frac{1}{E-\lambda} +\beta \delta (E-\lambda) \eeq and \beq \int_{-\infty}^{\infty}f_{\lambda}(E)dE=-\frac{(\alpha -\lambda )c_{\lambda}}{a} \eeq 

From this we find:  \beq -\rho a c_{\lambda} \int_{-\infty}^{\infty}P \frac {1}{E-\lambda}dE+\beta = - \frac{(\alpha - \lambda )c_{\lambda}}{a}\eeq 

The integral is zero and we find: 
\beq \beta =- \frac{(\alpha-\lambda)c_{\lambda}}{a} \eeq 

If we insert this we find the following expression for \( f_{\lambda}(E)\):
\beq f_{\lambda}(E)=-c_{\lambda}[\rho a P \frac{1}{E-\lambda}+\frac{\alpha -\lambda}{a} \delta(E-\lambda )] \eeq 

Except for a phase factor the coefficient \(c_{\lambda}\) can be calculated from the normalization condition. We have: 
\beq
\eqalign{
&\int_{-\infty}^{\infty}f_{\lambda}^{*}(E)f_{\lambda'}(E)dE\cr&=c_{\lambda}^{*}c_{\lambda'} \int_{-\infty}^{\infty}[\rho a P \frac{1}{E-\lambda}+\frac{\alpha - \lambda}{a}\delta(E-\lambda)]\cr&\qquad.[\rho a P \frac{1}{E-\lambda'}+\frac{\alpha-\lambda'}{a}\delta(E-\lambda')]dE\cr
&= c_{\lambda}^{*}c_{\lambda'}[\rho^{2}a^{2}\int_{-\infty}^{\infty}P\frac{1}{E-\lambda}.P\frac{1}{E-\lambda'}dE\cr
&\qquad+\frac{\alpha-\lambda}{a}\int_{-\infty}^{\infty}\delta (E-\lambda).\rho a P\frac{1}{E-\lambda'}dE\cr
&\qquad+\frac{\alpha-\lambda'}{a}\int_{-\infty}^{\infty}\delta (E-\lambda').\rho a P \frac{1}{E-\lambda}dE\cr
&\qquad+ \frac{(\alpha - \lambda)(\alpha - \lambda')}{a^{2}}\int_{-\infty}^{\infty}\delta (E-\lambda).\delta (E-\lambda')dE]
}
\eeq 
 
The first integral on the right is not zero. We have:
\beq
\eqalign{
&\int_{-\infty}^{\infty}P\frac{1}{E-\lambda}.P\frac{1}{E-\lambda'}dE\cr
&=\int_{-\infty}^{\infty}[\frac{1}{E-\lambda-i\epsilon}-i\pi\delta (E-\lambda)].[\frac{1}{E-\lambda'-i\epsilon}-i\pi\delta(E-\lambda')]dE\cr
&=\int_{-\infty}^{\infty}\frac{1}{(E-\lambda-i\epsilon)(E-\lambda'-i\epsilon)}dE-i\pi\int_{-\infty}^{\infty}\delta(E-\lambda)\frac{dE}{E-\lambda'-i\epsilon}\cr&\qquad-i\pi\int_{-\infty}^{\infty}\delta(E-\lambda')\frac{dE}{E-\lambda-i\epsilon}-\pi^{2}\delta(\lambda-\lambda') 
}
\eeq 

Here the first integral is zero as follows from contour integration. The next two terms give:  \beq -i\pi[\frac{1}{\lambda-\lambda'-i\epsilon}+\frac{1}{\lambda'-\lambda-i\epsilon}]=\frac{2\pi\epsilon}{(\lambda-\lambda')^{2}+\epsilon^{2}} \eeq 

We have: \beq \int_{-\infty}^{\infty}\frac{2\pi\epsilon}{(\lambda-\lambda')^{2}+\epsilon^{2}}d\lambda=2\pi \int_{-\infty}^{\infty}\frac{d\lambda}{1+\lambda^{2}}=2\pi^{2} \eeq 

Both terms together therefore give: 
\beq 2\pi^{2}\delta(\lambda-\lambda')\eeq 

We therefore find: 
\beq \int_{-\infty}^{\infty}P\frac{1}{E-\lambda}.P\frac{1}{E-\lambda'}dE=\pi^{2}\delta(\lambda-\lambda') \eeq 

The following two integrals together give: 
\beq
\eqalign{
&\frac{\alpha-\lambda}{a}\int_{-\infty}^{\infty}\delta(E-\lambda).\rho a P \frac{1}{E-\lambda'}dE\cr
&\qquad+\frac{\alpha-\lambda'}{a}\int_{-\infty}^{\infty}\delta(E-\lambda').\rho a P\frac{1}{E-\lambda}dE\cr
&= \rho(\alpha-\lambda)P\frac{1}{\lambda-\lambda'}+\rho(\alpha-\lambda') P \frac{1}{\lambda'-\lambda}=-\rho
}
\eeq 

The last integral is:
\beq 
\frac{(\alpha-\lambda)(\alpha-\lambda')}{a^{2}}\int_{-\infty}^{\infty}\delta(E-\lambda).\delta(E-\lambda')dE=\frac{(\alpha-\lambda)^{2}}{a^{2}}\delta(\lambda-\lambda') 
\eeq 

Everything taken together: 
\beq
\eqalign{
&\int_{-\infty}^{\infty}f_{\lambda}^{*}(e)f_{\lambda'}(E)dE\cr
&=c_{\lambda}^{*}c_{\lambda'}\int_{-\infty}^{\infty}[\rho a P\frac{1}{E-\lambda}+\frac{\alpha-\lambda}{a}\delta(E-\lambda)]\cr
&\qquad.[\rho a P\frac{1}{E-\lambda'}+\frac{\alpha-\lambda'}{a}\delta(E-\lambda')]dE\cr
&= c_{\lambda}^{*}c_{\lambda'}{\{}[\frac{(\alpha-\lambda)^{2}}{a^{2}}+\rho^{2}a^{2}\pi^{2}]\delta(\lambda-\lambda')-\rho{\}}
}
\eeq 

Thus we find:
\beq
\eqalign{
\langle\tilde{\lambda}|\tilde{\lambda'}\rangle&=\frac{1}{\rho}\int_{-\infty}^{\infty}f_{\lambda}^{*}(E)f_{\lambda'}(E)dE+c_{\lambda}^{*}c_{\lambda'}\cr
&= c_{\lambda}^{*}c_{\lambda'}[\frac{(\alpha-\lambda)^{2}}{\rho a^{2}}+\rho a^{2}\pi^{2}]\delta(\lambda-\lambda')=\delta(\lambda-\lambda')
}
\eeq 
and so we can make the following choice: \beq c_{\lambda}=\frac{-\sqrt{\rho}a}{\alpha-\lambda+i\pi\rho a^{2}} \eeq so that:  \beq f_{\lambda}(E)=\frac{-\sqrt{\rho}a}{\alpha-\lambda+i\pi\rho a^{2}}[\rho a P\frac{1}{E-\lambda}+\frac{\alpha-\lambda}{a}\delta(E-\lambda)] \eeq 

Finally we find the following exact expression: 
\beq |\tilde{\lambda}\rangle=\frac{\sqrt{\rho} a}{\alpha-\lambda+i\pi\rho a^{2}}[\int_{-\infty}^{\infty}\rho a P\frac{1}{E-\lambda}|E\rangle dE+\frac{\alpha-\lambda}{a}|\lambda\rangle-|a\rangle] \eeq 

The spectral representation of \(H\) becomes: 
\beq
\eqalign{
H&=\int_{-\infty}^{\infty}\lambda|\tilde{\lambda}\rangle\langle\tilde{\lambda}|d\lambda\cr
&= \int_{-\infty}^{\infty}\frac{\rho a^{2}\lambda d\lambda}{(\alpha-\lambda)^{2}+\pi^{2}\rho^{2} a^{4}}[\rho a \int_{-\infty}^{\infty}P\frac{1}{E-\lambda}|E\rangle dE+\frac{\alpha-\lambda}{a}|\lambda\rangle-|a\rangle]\cr
&\qquad.[\rho a \int_{-\infty}^{\infty}P\frac{1}{E'-\lambda}\langle E'|dE'+\frac{\alpha-\lambda}{a} \langle\lambda|-\langle a|]
}
\eeq 

This immediately leads to the following expression:
\beq
\eqalign{
e^{-\frac{it}{\hbar}H}&=\int_{-\infty}^{\infty}\frac{\rho a^{2} e^{-\frac{it}{\hbar}\lambda} d\lambda}{(\alpha-\lambda)^{2}+\pi^{2}\rho^{2}a^{4}}\cr
&\qquad.[\rho a \int_{-\infty}^{\infty}P\frac{1}{E-\lambda}|E\rangle dE+\frac{\alpha-\lambda}{a}|\lambda \rangle-|a\rangle]\cr
&\qquad.[\rho a \int_{-\infty}^{\infty}P\frac{1}{E'-\lambda}\langle E'|dE'+\frac{\alpha-\lambda}{a}\langle\lambda|-\langle a|]
}
\eeq 

From this it follows that:
\beq \langle a|e^{-\frac{it}{\hbar}H}|a\rangle=\int_{-\infty}^{\infty}\frac{\rho a^{2} e^{-\frac{it}{\hbar}\lambda}}{(\alpha-\lambda)^{2}+\pi^{2}\rho^{2} a^{4}}d\lambda \eeq 

This integral can be evaluated by contour integration. We have already assumed \(t\geq 0\), therefore, by closing the contour with a semicircle at infinity in the lower half plane we find:
\beq 
\eqalign{
&\int_{-\infty}^{\infty}\frac{\rho a^{2}e^{-\frac{it}{\hbar}\lambda}}{(\alpha-\lambda)^{2}+\pi^{2}\rho^{2}a^{4}}d\lambda\cr
&=-2\pi i\lim_{\lambda \rightarrow \alpha -i\pi\rho a^{2}} \frac{\rho a^{2} e^{-\frac{it}{\hbar}\lambda}(\lambda-\alpha+i\pi \rho a^{2})}{(\lambda-\alpha)^{2}+\pi^{2}\rho^{2} a^{4}}\cr
&= -2\pi i \lim_{\lambda-\alpha-i\pi\rho a^{2}}\frac{\rho a^{2}e^{-\frac{it}{\hbar}\lambda}}{\lambda-\alpha-i\pi\rho a^{2}}\cr
&=-2\pi i \frac{\rho a^{2} e^{-\frac{it}{\hbar}(\alpha-i\pi\rho a^{2})}}{-2i\pi\rho a^{2}}=e^{-\frac{it}{\hbar}(\alpha-i\pi\rho a^{2})}
}
\eeq 

Thus we find: \beq \langle a|e^{-\frac{it}{\hbar}H}|a\rangle=e^{-\frac{it}{\hbar}(\alpha-i\pi\rho a^{2})} \eeq 

In the course of time the probability of finding the original state back is equal to: \beq W_{a}(t)=|\langle a|e^{-\frac{it}{\hbar}H}|a\rangle|^{2}=e^{-\frac{2\pi\rho a^{2}}{\hbar}t} \eeq 

This is the exponential law of radioactive decay. Note that the coefficient of \( t \) in the exponent is in agreement with Fermi's Golden Rule.

Except for the simplified model specifications no approximation is made. Striking is that for {\it negative} \(t\) the same integral over \(\lambda\) generates a plus sign in the exponent, so that time symmetry is restored. There is apparently no question of time irreversibility. A simple kink in the time curve appears (see Figure~1).


\begin{figure}
\epsfig{file=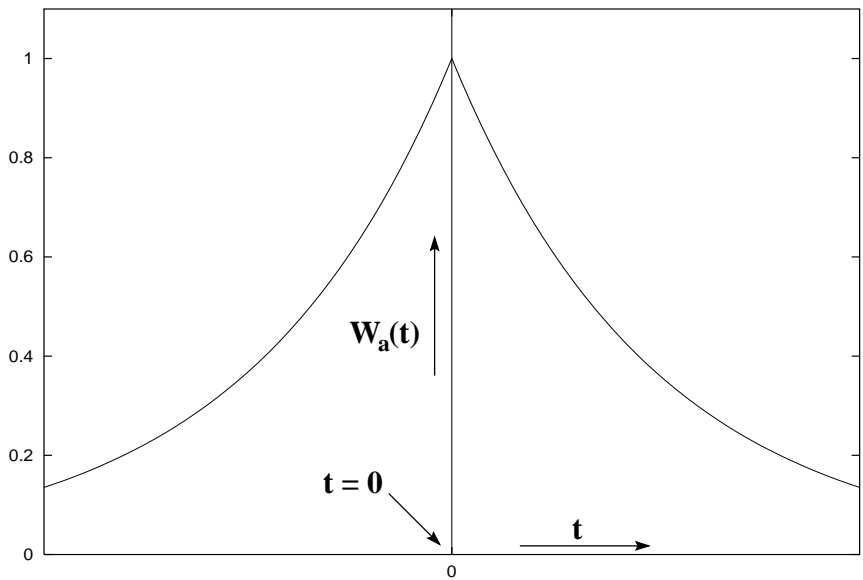,width=\textwidth}
\caption{Formal survival rate as function of time.}
\end{figure}

\section{Extension of the model}
In Section 2 we assumed \(\rho\) and \(a\) to be independent of E. We now introduce E-dependence. We then have: \beq H_{0}|E\rangle=E|E\rangle,\hspace{1cm}H_{0}|a\rangle=\alpha|a\rangle,\hspace{1cm}-\infty < E< \infty \eeq \beq \langle E|E'\rangle =\frac{1}{\rho(E)}\delta (E-E'),\hspace{1cm}\langle E|a\rangle=\langle a|E\rangle=0,\hspace{1cm}\langle a|a \rangle=1 \eeq \beq \int_{-\infty}^{\infty}\rho(E)|E\rangle \langle E|dE+|a\rangle \langle a|=I,\hspace{1cm}\rho>0 \eeq  and  \beq H'|E\rangle=a(E)|a\rangle,\hspace{1cm}H'|a\rangle=\int_{-\infty}^{\infty}\rho(E)a(E)|E\rangle dE \eeq 

Again we define: \beq |\tilde{\lambda}\rangle=\int_{-\infty}^{\infty}f_{\lambda}(E)|E\rangle dE+c_{\lambda}|a\rangle \eeq  

and try to solve the eigenvalue equation: \beq (H-\lambda)|\tilde{\lambda}\rangle=(H_{0}+H'-\lambda)|\tilde{\lambda}\rangle=0 \eeq with the normalization condition: \beq \langle \tilde{\lambda}|\tilde{\lambda'}\rangle=\int_{-\infty}^{\infty}\frac{1}{\rho(E)}f_{\lambda}^{*}(E)f_{\lambda'}(E)dE+c_{\lambda}^{*}c_{\lambda'}=\delta(\lambda-\lambda')\eeq 

This now gives:
\beq 
\eqalign{
(H_{0}+H'-\lambda)|\tilde{\lambda}\rangle&=(H_{0}+H'-\lambda)[\int_{-\infty}^{\infty}f_{\lambda}(E)|E\rangle dE+c_{\lambda}|a\rangle ]\cr
&= \int_{-\infty}^{\infty}f_{\lambda}(E)[(E-\lambda)|E\rangle+a(E)|a\rangle ]dE\cr
&\qquad+c_{\lambda}[(\alpha-\lambda)|a\rangle+\int_{-\infty}^{\infty}\rho(E)a(E)|E\rangle dE]=0 
}
\eeq 

This leads to: \beq (E-\lambda)f_{\lambda}(E)+\rho(E)a(E)c_{\lambda}=0 \eeq and \beq (\alpha -\lambda)c_{\lambda}+\int_{-\infty}^{\infty}a(E)f_{\lambda}(E)dE=0 \eeq 

So we find: \beq f_{\lambda}(E)=-\rho(E)a(E)c_{\lambda}P\frac{1}{E-\lambda}+\beta \delta(E-\lambda) \eeq and \beq -c_{\lambda}\int_{-\infty}^{\infty}\rho(E) a^{2}(E)P\frac{1}{E-\lambda}dE+a(\lambda)\beta=-(\alpha - \lambda)c_{\lambda} \eeq 

Now the integral is not zero. We find instead what is called the Hilbert transform\footnote{Erd\'{e}lyi, Bateman Manuscript Project, "Tables of Integral Transforms II", page 243. (McGraw Hill, 1954)} of the function \(\rho(E)a^{2}(E)\).

Define: \beq \eta(E)=\rho(E)a^{2}(E) \eeq 

It is assumed that \(\eta(E)\rightarrow 0\) sufficiently fast when \(E\rightarrow \pm\infty\).

The Hilbert transform \(\sigma(\lambda)\) of \(\eta(E)\) is defined as: \beq \sigma(\lambda)=\frac{1}{\pi}\int_{-\infty}^{\infty}\eta(E)P\frac{1}{E-\lambda}dE \eeq 

In terms of this function we then have: \beq \beta=-c_{\lambda}\frac{\alpha-\lambda-\pi\sigma(\lambda)}{a(\lambda)} \eeq  so we find: \beq f_{\lambda}(E)=-c_{\lambda}[\rho(E)a(E)P\frac{1}{E-\lambda}+\frac{\alpha-E-\pi\sigma(E)}{a(E)}\delta(E-\lambda)] \eeq 

In order to make use of the normalization condition we evaluate:
\beq 
\eqalign{
\int_{-\infty}^{\infty}\frac{1}{\rho(E)}&f_{\lambda}^{*}(E)f_{\lambda'}(E)\cr
&=c_{\lambda}^{*}c_{\lambda'}\int_{-\infty}^{\infty}\frac{1}{\rho(E)}[\rho(E)a(E)P\frac{1}{E-\lambda}+\frac{\alpha-E-\pi\sigma(E)}{a(E)}\delta(E-\lambda)]\cr
&\qquad.[\rho(E)a(E)P\frac{1}{E-\lambda'}+\frac{\alpha-E-\pi\sigma(E)}{a(E)}\delta(E-\lambda')]dE
}
\eeq 

The integral: \beq \int_{-\infty}^{\infty}\rho(E) a^{2}(E)P\frac{1}{E-\lambda}P\frac{1}{E-\lambda'}dE=\int_{-\infty}^{\infty}\eta(E)P\frac{1}{E-\lambda}P\frac{1}{E-\lambda'}dE \eeq 

can be determined and gives (see the Appendix): \beq \pi\frac{\sigma(\lambda)-\sigma(\lambda')}{\lambda-\lambda'}+\pi^{2}\eta(\lambda)\delta(\lambda-\lambda') \eeq 

The remaining terms give no problems and we end up with: 
\beq
\eqalign{
\langle \tilde{\lambda}|\tilde{\lambda'}\rangle&=\int_{-\infty}^{\infty}\frac{1}{\rho(E)}f_{\lambda}^{*}(E)f_{\lambda'}(E)dE+c_{\lambda}^{*}c_{\lambda'}\cr
&= c_{\lambda}^{*}c_{\lambda'}{\{}\frac{1}{\eta(\lambda)}[\alpha-\lambda-\pi\sigma(\lambda)]^{2}+\eta(\lambda){\}}\delta(\lambda-\lambda')=\delta(\lambda-\lambda')
}
\eeq 

Apparently we can choose: \beq c_{\lambda}=\frac{-\sqrt{\eta(\lambda)}}{\alpha-\lambda-\pi\sigma(\lambda)+i\pi\eta(\lambda)} \eeq 

This gives:
\beq 
\eqalign{
|\tilde{\lambda}\rangle&=\frac{\sqrt{\eta(\lambda)}}{\alpha-\lambda-\pi\sigma(\lambda)+i\pi\eta(\lambda)}[\int_{-\infty}^{\infty}\rho(E)a(E)P\frac{1}{E-\lambda}|E\rangle dE\cr
&\qquad+ \frac{\alpha-\lambda-\pi\sigma(\lambda)}{a(\lambda)}|\lambda\rangle-|a\rangle]
}
\eeq 

From this it follows that:\beq \langle a|\tilde{\lambda}\rangle =\frac{-\sqrt{\eta(\lambda)}}{\alpha-\lambda-\pi\sigma(\lambda)+i\pi\eta(\lambda)} \eeq  

and thus we obtain:
\beq
\eqalign{
\langle a|e^{-\frac{it}{\hbar}H}|a\rangle&=\int_{-\infty}^{\infty}e^{-\frac{it}{\hbar}\lambda}\langle a|\tilde{\lambda}\rangle \langle\tilde{\lambda}|a\rangle d\lambda \cr
&= \int_{-\infty}^{\infty}\frac{\eta(\lambda)e^{-\frac{it}{\hbar}\lambda}d\lambda}{[\alpha-\lambda-\pi\sigma(\lambda)]^{2}+\pi^{2}\eta^{2}(\lambda)}
}
\eeq 

This is again an exact expression. If \(a\) is small also \(\sigma\) is small and a good approximation can be obtained by writing:
\beq 
\eqalign{
\langle a|e^{-\frac{it}{\hbar}H}|a\rangle&\approx \int_{-\infty}^{\infty}\frac{\eta(\alpha)e^{-\frac{it}{\hbar}\lambda}d\lambda}{[\alpha-\lambda-\pi\sigma(\alpha)]^{2}+\pi^{2}\eta^{2}(\alpha)}\cr
&=e^{-\frac{it}{\hbar}[\alpha-\pi\sigma(\alpha)-i\pi\eta(\alpha)]}
}
\eeq

Note that this is again in agreement with Fermi's Golden Rule.

Let us now consider the exact integral expression. By making use of theorems on Hilbert transforms we can get insight into the analytical properties of the integrand. The relevant theorems are proven in the Appendix and sound:

 1. If \(\sigma(y)\) is the Hilbert transform of \(\eta(x)\): \beq \sigma(y)=\frac{1}{\pi}\int_{-\infty}^{\infty}\eta(x)P\frac{1}{x-y}dx \eeq then \(-\eta(y)\) is the Hilbert transform of \(\sigma(x)\): \beq \eta(y)=-\frac{1}{\pi}\int_{-\infty}^{\infty}\sigma(x)P\frac{1}{x-y}dx \eeq 

 2. The function: \beq \xi(z)\stackrel{\rm def}{=}\frac{1}{2\pi i}\int_{-\infty}^{\infty}\frac{\sigma(x)+i\eta(x)}{x-z}dx \eeq 

is zero in the lower half \(z\)-plane, analytic and regular in the upper half \(z\)-plane and has the property: \beq \lim_{z\downarrow x_{0}}\xi(z)=\sigma(x_{0})+i\eta(x_{0}) \eeq 

Here \(x\) and \(x_{0}\) are points on the real axis. The function can be analytically continued from the upper to the lower half \(z\)-plane but may not be regular there. It is clear that \(\xi^{*}(z)\) has just the opposite properties.

 3. The function:\beq \tilde{\sigma} (x,y)=\frac{1}{\pi}\int_{-\infty}^{\infty}\eta(z)P\frac{1}{z-x}P\frac{1}{z-y}dz \eeq 

satisfies the property: \beq \tilde{\sigma} (x,y)=\frac{\sigma(x)-\sigma(y)}{x-y}+\pi\eta(x)\delta(x-y) \eeq

The latter theorem has been used before and will be used later on.

We can now rewrite the matrix element in terms of \(\xi\), \(\xi^{*}\) and \(\eta\): \beq \langle a|e^{-\frac{it}{\hbar}H}|a\rangle=\int_{-\infty}^{\infty}\frac{\eta(\lambda)e^{-\frac{it}{\hbar}\lambda}d\lambda}{[\alpha-\lambda-\pi\xi(\lambda)].[\alpha-\lambda-\pi\xi^{*}(\lambda)]} \eeq 

If \(\eta(\lambda)\) is regular in the lower half \(\lambda\)-plane except for poles then since:
\beq
\eta(\lambda)=\frac{\xi(\lambda)-\xi^{*}(\lambda)}{2i}
\eeq  and \(\xi^{*}(\lambda)\) is regular in the lower half plane, also (the analytical continuation of) \(\xi(\lambda)\) is regular in the lower half plane except for poles. For \(t\neq0\) we can split the integral into two parts: 

\beq
\eqalign{
&\int_{-\infty}^{\infty}\frac{\eta(\lambda)e^{-\frac{it}{\hbar}\lambda}d\lambda}{[\alpha-\lambda-\pi\xi(\lambda)].[\alpha-\lambda-\pi\xi^{*}(\lambda)]}\cr
&= \frac{1}{2\pi i}[\int_{-\infty}^{\infty}\frac{e^{-\frac{it}{\hbar}\lambda}d\lambda}{\alpha-\lambda-\pi\xi(\lambda)}-\int_{-\infty}^{\infty}\frac{e^{-\frac{it}{\hbar}\lambda}d\lambda}{\alpha-\lambda-\pi\xi^{*}(\lambda)}]
}
\eeq 

In order to evaluate this expression we have to solve the equation: \beq \alpha-\lambda-\pi\xi(\lambda)=0 \eeq 

It is interesting to see what happens when we introduce a scaling factor and replace \(\eta\) by \(\tau\eta\)\hspace{0.3cm}\((\tau>0)\). Then \(\xi\) is replaced by \(\tau\xi\). We then have to solve the equation: \beq \alpha-\lambda-\pi\tau\xi(\lambda)=0 \eeq 

When \(\tau\) approaches zero one of the roots approaches \(\alpha\). Then in first approximation we have: \beq \alpha-\lambda-\pi\tau\xi(\alpha)=0 \eeq and we find: \beq \lambda_{0} \approx \alpha-\pi\tau\xi(\alpha)=\alpha-\pi\tau[\sigma(\alpha)+i\eta(\alpha)] \eeq 

This is what we have seen before. Since \(\eta(\alpha)>0\) this zero lies in the lower half plane and since this root never becomes real for any value of \(\tau\) it stays in the lower half plane for all values of \(\tau\).

The other roots approach poles in \(\xi(\lambda)\). Let \(\lambda'\) be such a pole. Then, near \(\lambda'\), we have: \beq \xi(\lambda)\approx\frac{r_{\lambda'}}{\lambda-\lambda'} \eeq 

and we find: \beq \lambda_{0}\approx\lambda'+\pi\tau\frac{r_{\lambda'}}{\alpha-\lambda'} \eeq 

Since \(\xi(\lambda)\) is regular in the upper half plane, for sufficiently small \(\tau\) these roots are all located in the lower half plane and since the roots are never real for any value of \(\tau\) they remain in the lower half plane for all values of \(\tau\). Meanwhile, \(\xi^{*}(\lambda)\) is regular in the lower half plane, so for sufficiently small \(\tau\) there cannot be solutions of the equation: \beq \alpha-\lambda-\pi\tau\xi^{*}(\lambda)=0 \eeq 

located in the lower half plane. Now when we let \(\tau\) move from infinitesimal to regular values zero's will not move in and cannot spontaneously be created so for any \(\tau\) there will be no solutions of the equation in the lower half plane. For \(t>0\) the conclusion is that: \beq \int_{-\infty}^{\infty}\frac{e^{-\frac{it}{\hbar}\lambda}d\lambda}{\alpha-\lambda-\pi\xi^{*}(\lambda)}=0 \eeq and we find: \beq \langle a|e^{-\frac{it}{\hbar}H}|a\rangle=\frac{1}{2\pi i}\int_{-\infty}^{\infty}\frac{ e^{-\frac{it}{\hbar}\lambda}d\lambda}{\alpha-\lambda-\pi\xi(\lambda)} \eeq 

Let \(\lambda_{0}\) be a root. Then \(\Im\lambda_{0}<0\) and its contribution to the integral is: \beq -\frac{1}{\pi\xi'(\lambda_{0})+1}e^{-\frac{it}{\hbar}\lambda_{0}} \eeq 

We end up with the following expression for \(t\geq 0\): \beq \langle a|e^{-\frac{it}{\hbar}H}|a\rangle=\sum_{\lambda_{0}}\gamma_{\lambda_{0}}e^{-\frac{it}{\hbar}\Re\lambda_{0}}e^{-\frac{t}{\hbar}|\Im \lambda_{0}|} \eeq where the \(\gamma\) satisfy the necessary but not sufficient condition: \beq \sum_{\lambda_{0}}\gamma_{\lambda_{0}}=1 \eeq 

The result is a sum of exponentials decreasing with time. One of them ultimately will dominate. Note that the time derivative of \beq W_{a}(t)=|\langle a|e^{-\frac{it}{\hbar}H}|a\rangle|^{2} \eeq 

in the limit \(t\downarrow 0\) is not zero. Still there is time reversal symmetry. Therefore again a kink appears at \(t=0\) and the time derivative at \(t=0\) does not exist.

\section{The influence of the energy lower bound}

Without loss of generality we may assume that \(E=0\) is the lowest energy value of the free Hamiltonian \(H_{0}\). In that case we have, with \(\alpha\) assumed \(\neq 0\): \beq H_{0}|E\rangle=E|E\rangle,\hspace{1cm} H_{0}|a\rangle=\alpha|a\rangle,\hspace{1cm}0<E<\infty \eeq \beq \langle E|E'\rangle=\frac{1}{\rho(E)}\delta(E-E'),\hspace{1cm}\langle E|a\rangle=\langle a|E\rangle=0,\hspace{1cm}\langle a|a\rangle=1 \eeq \beq \int_{0}^{\infty}\rho(E)|E\rangle\langle E|dE+|a\rangle\langle a|=I,\hspace{1cm} \rho>0 \eeq and \beq H'|E\rangle=a(E)|a\rangle,\hspace{1cm}H'|a\rangle=\int_{0}^{\infty}\rho(E)a(E)|E\rangle dE \eeq 

As before we consider normalized eigenkets of the Hamiltonian \(H=H_{0}+H'\). We put \(H\) in its spectral form: \beq H=\int_{-\infty}^{\infty}\lambda|\tilde{\lambda}\rangle\langle\tilde{\lambda}|d\lambda \eeq \beq H|\tilde{\lambda}\rangle=\lambda|\tilde{\lambda}\rangle \eeq \beq \langle\tilde{\lambda}|\tilde{\lambda}'\rangle=\delta(\lambda-\lambda') \eeq \beq \int_{-\infty}^{\infty}|\tilde{\lambda}\rangle\langle\tilde{\lambda}|d\lambda=I \eeq 

Note here that we do not assume a lower bound on the eigenvalues. We now write:  \beq |\tilde{\lambda}\rangle=\int_{0}^{\infty}f_{\lambda}(E)|E\rangle dE+c_{\lambda}|a\rangle \eeq 

The eigenvalue equation gives:
\beq 
\eqalign{
(H_{0}+H'-\lambda)|\tilde{\lambda}\rangle&=(H_{0}+H'-\lambda)[\int_{0}^{\infty}f_{\lambda}(E)|E\rangle dE+c_{\lambda}|a\rangle]\cr
&= \int_{0}^{\infty}f_{\lambda}(E)[(E-\lambda)|E\rangle+a(E)|a\rangle]dE\cr
&\qquad+ c_{\lambda}[(\alpha-\lambda)|a\rangle+\int_{0}^{\infty}\rho(E)a(E)|E\rangle dE]=0
}
\eeq from which it follows that:
\beq (E-\lambda)f_{\lambda}(E)+\rho(E)a(E)c_{\lambda}=0
\eeq 

and \beq (\alpha-\lambda)c_{\lambda}+\int_{0}^{\infty}a(E)f_{\lambda}(E)dE=0 \eeq 

This gives, as before:\beq f_{\lambda}(E)=-\rho(E)a(E)c_{\lambda}P\frac{1}{E-\lambda}+\beta\delta(E-\lambda) \eeq 

When \(\lambda>0\) we obtain an equation for \(\beta\):\beq -c_{\lambda}\int_{0}^{\infty}\eta(E)P\frac{1}{E-\lambda}dE+a(\lambda)\beta=-(\alpha-\lambda)c_{\lambda} \eeq which can be solved: \beq \beta=c_{\lambda}\frac{\int_{0}^{\infty}\eta(E)P\frac{1}{E-\lambda}dE-(\alpha-\lambda)}{a(\lambda)} \eeq 

In that case we have: \beq f_{\lambda}(E)=-c_{\lambda}[\rho(E)a(E)P\frac{1}{E-\lambda}+\frac{\alpha-E-\pi\bar{\sigma}(E)}{a(E)}\delta(E-\lambda)] \eeq where: \beq \bar{\sigma}(\lambda)=\frac{1}{\pi}\int_{0}^{\infty}\eta(E)P\frac{1}{E-\lambda}dE=\frac{1}{\pi}\int_{-\infty}^{\infty}\eta(E)\theta(E)P\frac{1}{E-\lambda}dE \eeq 

and is therefore the Hilbert transform of the function \(\eta(E)\theta(E)\). We define: \beq \bar{\eta}(E)=\eta(E)\theta(E) \eeq 

We get the following expression for the eigenkets of H:
\beq
\eqalign{
|\tilde{\lambda}\rangle&=-c_{\lambda}\int_{0}^{\infty}\frac{1}{a(E)}{\{}\eta(E)P\frac{1}{E-\lambda}\cr
&\qquad+ [\alpha-E-\pi\bar{\sigma}(E)]\delta(E-\lambda){\}}|E\rangle dE+c_{\lambda}|a\rangle
}
\eeq 

We evaluate:
\beq 
\eqalign{
&\int_{0}^{\infty}\frac{1}{\rho(E)}f_{\lambda}^{*}(E)f_{\lambda'}(E)dE\cr
&= c_{\lambda}^{*}c_{\lambda'}\int_{0}^{\infty}\frac{1}{\eta(E)}{\{}\eta(E)P\frac{1}{E-\lambda}+[\alpha-E-\pi\bar{\sigma}(E)]\delta(E-\lambda){\}}\cr
&\qquad.{\{}\eta(E)P\frac{1}{E-\lambda'}+[\alpha-E-\pi\bar{\sigma}(E)]\delta(E-\lambda'){\}}dE
}
\eeq 

Here we have:
\beq 
\eqalign{
&\int_{0}^{\infty}\eta(E)P\frac{1}{E-\lambda}P\frac{1}{E-\lambda'}dE=\int_{-\infty}^{\infty}\bar{\eta}(E)P\frac{1}{E-\lambda}P\frac{1}{E-\lambda'}dE\cr
&\qquad= \pi\frac{\bar{\sigma}(\lambda)-\bar{\sigma}(\lambda')}{\lambda-\lambda'}+\pi^{2}\bar{\eta}(\lambda)\delta(\lambda-\lambda')
}
\eeq 

Again the remaining terms can immediately be evaluated and the result becomes:
\beq
\eqalign{
\langle \tilde{\lambda}|\tilde{\lambda'}\rangle&=\int_{0}^{\infty}\frac{1}{\rho(E)}f_{\lambda}^{*}(E)f_{\lambda'}(E)dE+c_{\lambda}^{*}c_{\lambda'}\cr
&= c_{\lambda}^{*}c_{\lambda'}{\{}\frac{1}{\eta(\lambda)}[\alpha-\lambda-\pi\bar{\sigma}(\lambda)]^{2}+\bar{\eta}(\lambda){\}}\delta(\lambda-\lambda')=\delta(\lambda-\lambda')
}
\eeq 

for \(\lambda, \lambda'>0\). This allows the normalization constant to be determined except for a phase. We choose: \beq c_{\lambda}=\frac{-\sqrt{\bar{\eta}(\lambda)}}{\alpha-\lambda-\pi\bar{\sigma}(\lambda)+i\pi\bar{\eta}(\lambda)}=\frac{-\sqrt{\bar{\eta}(\lambda)}}{\alpha-\lambda-\pi\bar{\xi}^{*}(\lambda)} \eeq 

Note the difference with the former expression. Moreover it is only valid for \(\lambda>0\).

Finally we obtain for positive \(\lambda\):
\beq
\eqalign{
|\tilde{\lambda}\rangle&=\frac{\sqrt{\bar{\eta}(\lambda)}}{\alpha-\lambda-\pi\bar{\xi}^{*}(\lambda)}[\int_{0}^{\infty}\frac{1}{a(E)}\eta(E)P\frac{1}{E-\lambda}|E\rangle dE\cr
&\qquad+\frac{\alpha-\lambda-\pi\bar{\sigma}(\lambda)}{a(\lambda)}|\lambda\rangle-|a\rangle]
}
\eeq 

and so we get: \beq \langle a|\tilde{\lambda}\rangle=\frac{-\sqrt{\bar{\eta}(\lambda)}}{\alpha-\lambda-\pi\bar{\xi}^{*}(\lambda)} \eeq 

When \(\lambda<0\) it is not possible to solve for \(\beta\) because then \(\delta(E-\lambda)\) is always zero. In that case we have: \beq f_{\lambda}(E)=-\frac{\rho(E)a(E)}{E-\lambda}c_{\lambda} \eeq 

When this is inserted we obtain: \beq \int_{0}^{\infty}\frac{\rho(E)a^{2}(E)}{E-\lambda}dE=\int_{0}^{\infty}\frac{\eta(E)}{E-\lambda}dE=\alpha-\lambda \eeq 

which gives: \beq \alpha-\lambda-\pi\bar{\sigma}(\lambda)=0 \eeq 

This is an equation for \(\lambda\) to be solved. We have for \(\lambda<0\) that the integral and all its derivatives are positive so for any \(\alpha\) with the property:  \beq \alpha<\lim_{\lambda\uparrow 0}\bar{\sigma}(\lambda) \eeq 

there is one and only one solution. If \(\alpha\) is larger then there is no solution. It might be that this limit is positive infinite. Then there is always one solution.

The solution is to be interpreted as a bound state. Let \(\lambda_{0}\) be this solution. We have for \(\lambda<0\): \beq |\tilde{\lambda}\rangle=-c_{\lambda}\int_{0}^{\infty}\frac{1}{a(E)}\frac{\eta(E)}{E-\lambda}|E\rangle dE+c_{\lambda}|a\rangle \eeq 

This gives:
\beq
\eqalign{
\langle\tilde{\lambda}|\tilde{\lambda}'\rangle&=c_{\lambda}^{*}c_{\lambda'}[\int_{0}^{\infty}\frac{\eta(E)dE}{(E-\lambda)(E-\lambda')}+1]\cr
&=c_{\lambda}^{*}c_{\lambda'}[\int_{-\infty}^{\infty}\bar{\eta}(E)P\frac{1}{E-\lambda}P\frac{1}{E-\lambda'}dE+1]\cr
&= c_{\lambda}^{*}c_{\lambda'}[\pi\frac{\bar{\sigma}(\lambda)-\bar{\sigma}(\lambda')}{\lambda-\lambda'}+\pi^{2}\bar{\eta}(\lambda)\delta(\lambda-\lambda')+1]\cr
&= c_{\lambda}^{*}c_{\lambda'}[\pi\frac{\bar{\sigma}(\lambda)-\bar{\sigma}(\lambda')}{\lambda-\lambda'}+1]
}
\eeq 

From this it follows that: \beq \langle\tilde{\lambda}_{0}|\tilde{\lambda}_{0}\rangle=c_{\lambda_{0}}^{*}c_{\lambda_{0}}[\pi\bar{\sigma}'(\lambda_{0})+1]=1 \eeq 
and thus we have: \beq c_{\lambda_{0}}^{*}c_{\lambda_{0}}=\frac{1}{\pi\bar{\sigma}'(\lambda_{0})+1} \eeq 

where: \beq \bar{\sigma}'(\lambda_{0})=\frac{1}{\pi}\int_{0}^{\infty}\frac{\eta(E)}{(E-\lambda_{0})^{2}}dE>0 \eeq 

The following expression results if there is a bound state:
\beq
\eqalign{
\langle a|e^{-\frac{it}{\hbar}\lambda}|a\rangle&=\int_{0}^{\infty}e^{-\frac{it}{\hbar}\lambda}\langle a|\tilde{\lambda}\rangle\langle\tilde{\lambda}|a\rangle d\lambda+e^{-\frac{it}{\hbar}\lambda_{0}}\langle a|\tilde{\lambda}_{0}\rangle\langle\tilde{\lambda}_{0}|a\rangle\cr
&= \int_{0}^{\infty}\frac{\bar{\eta}(\lambda)e^{-\frac{it}{\hbar}\lambda}d\lambda}{[\alpha-\lambda-\pi\bar{\sigma}(\lambda)][\alpha-\lambda-\pi\bar{\sigma}^{*}(\lambda)]}+\frac{e^{-\frac{it}{\hbar}\lambda_{0}}}{\pi\bar{\sigma}'(\lambda_{0})+1}
}
\eeq 

The second term on the right hand side oscillates forever. That means that when one prepares the state \(|a\rangle\) this state may contain a contribution from a bound state which does not decay. The first term can be rewritten in the form: \beq \frac{1}{2\pi i}[\int_{0}^{\infty}\frac{e^{-\frac{it}{\hbar}\lambda}d\lambda}{\alpha-\lambda-\pi\bar{\xi}(\lambda)}-\int_{0}^{\infty}\frac{e^{-\frac{it}{\hbar}\lambda}d\lambda}{\alpha-\lambda-\pi\bar{\xi}^{*}(\lambda)}] \eeq 

Since the integrands have a possible pole on the negative real axis, extension of the integration path to minus infinity is not immediately possible. We have to avoid this pole and we do that by choosing the path of integration for the first integral as shown in Figure~2.

\begin{figure}[htbp]
  \centerline{\epsfig{file=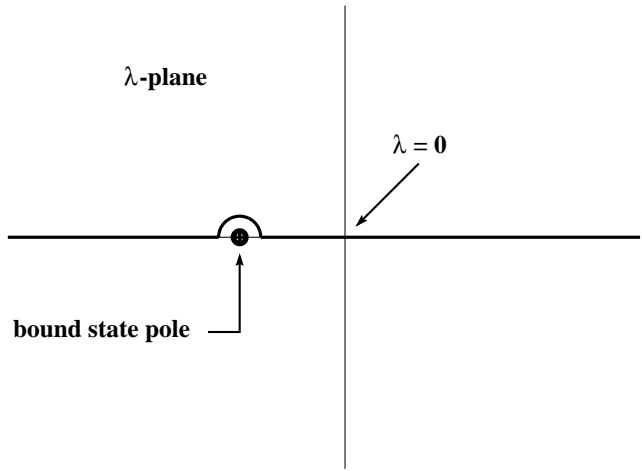,width=8.5cm}}
  \caption{Integration path of the first integral.}
\end{figure}


If the same integration path were chosen for the second integral the extension of the integration path to minus infinity would not have changed the results, because the integrands are equal. However, then the second integral would not be zero. This integral is only zero when the integration path be chosen as in Figure~3. 
\begin{figure}[htbp]
  \centerline{\epsfig{file=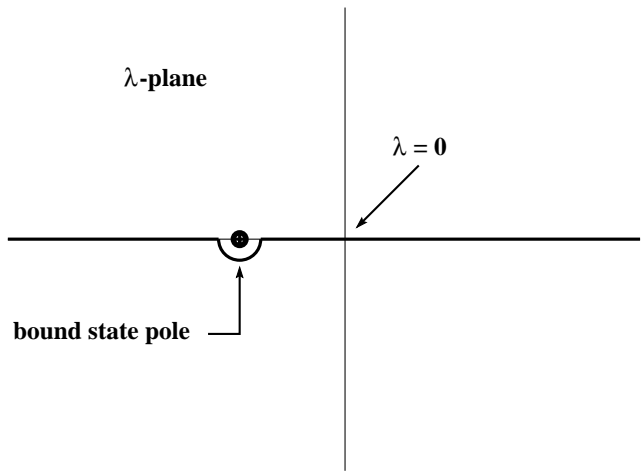,width=8.5cm}}
  \caption{Integration path of the second integral.}
\end{figure}


The difference is just the contribution from the bound state pole and thus we find that: \beq \langle a|e^{-\frac{it}{\hbar}H}|a\rangle=\frac{1}{2\pi i}\int\frac{e^{-\frac{it}{\hbar}\lambda}d\lambda}{\alpha-\lambda-\pi\bar{\xi}(\lambda)} \eeq 

with the path of integration taken along the entire real axis except for the bound state pole and which is passed through the upper half plane, i.e. the contour is that of Figure 2. The integrand is analytic and regular in the upper half plane (on this particular Riemann sheet) except for the pole on the negative real axis.

The right hand side of this expression causes trouble because the functions \(\bar{\eta}(\lambda)\) and \(\bar{\sigma}(\lambda)\) are definitely nonanalytic in the lower half plane and a contour integration in the way used above is not possible. Suppose that \(\eta(\lambda)\) be analytic and regular except for poles; \(\eta(\lambda)>0\) and finite for \(\lambda>0\); \(\eta(0)\) finite. Then \(\eta(\lambda)\theta(\lambda)\) is "analytic" in the sense that it has a branch cut which on both sides of the real axis stretches itself out towards infinity and passes through the origin. The cut separates two analytic functions, one of them being identically zero, the other being \(\eta(\lambda)\). By using a similar argument as before one can now easily prove that \(\bar{\xi}(\lambda) \) is still regular in the upper half plane and that its analytic continuation is regular in the lower half plane except for poles and a branch cut starting at the origin. Now the point \(\lambda=0\) is necessarily a true branch point. We take the cut along the negative imaginary axis. Except for the bound state pole all poles are on the right hand side of the cut and in the lower half plane.

In order to evaluate the integral we deform the contour as to wrap the cut: The right hand side is rotated clockwise and the left hand side anticlockwise towards the negative imaginary axis. In the mean time poles are passed which give exponential contributions to the integral. The problem now is to evaluate the remaining contour integral. In Figure~4 the branch cut, possible poles and the original integration path are drawn. Note how the branch point is evaded.

\begin{figure}[htbp]
  \centerline{\epsfig{file=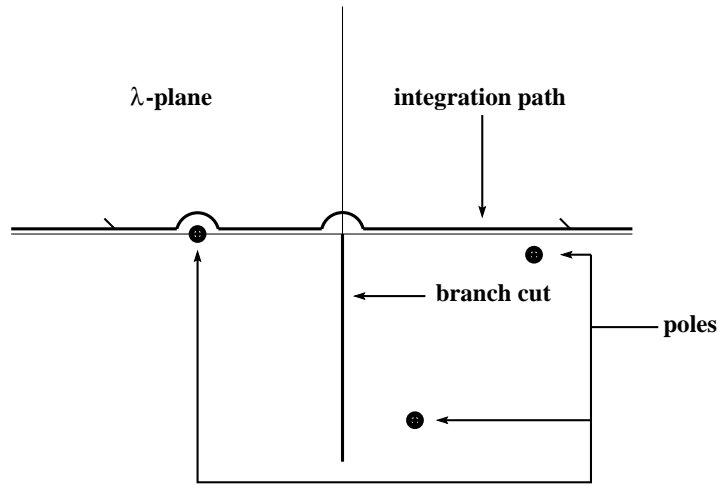,width=9.5cm}}
  \caption{Intermediate integration path.}
\end{figure}


After deformation of the contour the situation is as sketched in Figure~5.

\begin{figure}[htbp]
  \centerline{\epsfig{file=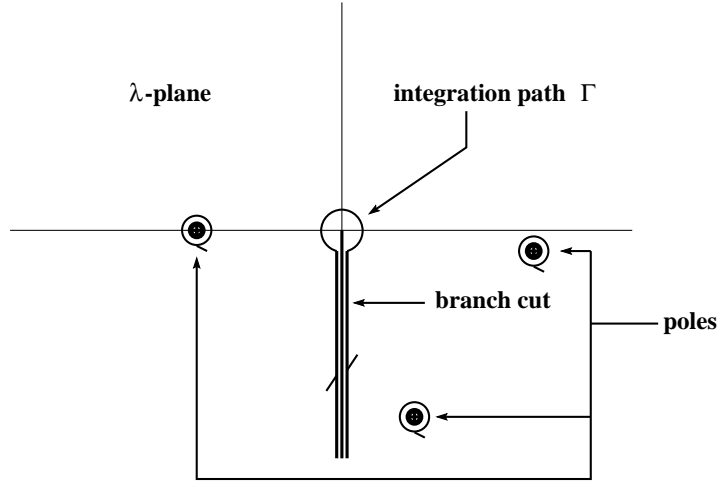,width=9.5cm}}
  \caption{Final integration path.}
\end{figure}


Let \(\Gamma \) be the final contour. The poles on the right hand side of the imaginary axis give damped oscillatory contributions, the pole on the left hand side gives a pure oscillatory contribution from the bound state.

The remaining contour integral is:
\beq
\eqalign{
\frac{1}{2\pi i}\int_{\Gamma}\frac{e^{-\frac{it}{\hbar}\lambda}d\lambda}{\alpha-\lambda-\pi\bar{\xi}(\lambda)}&=\frac{1}{2\pi i}\int_{0}^{-i\infty}F(\lambda)e^{-\frac{it}{\hbar}\lambda}d\lambda\cr
&=-\frac{1}{2\pi}\int_{0}^{\infty}F(-i\lambda)e^{-\frac{t}{\hbar}\lambda}d\lambda
}
\eeq 

where \(F(\lambda)\) is the "jump across the cut" of the function: \beq \frac{1}{\alpha-\lambda-\pi\bar{\xi}(\lambda)} \eeq

The integral is apparently a linear combination of an infinite number of decaying exponentials. The function \(F(\lambda)\) cannot be identically equal to zero between \(0\) and some point on the negative imaginary axis, otherwise \(\lambda=0\) is not a true branch point. We find therefore that for \(t\rightarrow\infty\) the integral will dominate any exponential of the type \(e^{-\frac{t}{\tau}}\) for \(\tau>0\). The conclusion is therefore that there will be a non-exponential contribution to the transition matrix element which will ultimately dominate any exponential decay. This is the anomaly.

Let us now discuss some details of this anomaly. Let \(\eta(\lambda)\) be a real, rational analytic function of \(\lambda\), positive when \(\lambda\) is positive. Then for {\it negative} \(\lambda\) we find that: \beq \bar{\sigma}(\lambda)=-\frac{1}{\pi}\eta(\lambda)\ln(-\frac{\lambda}{c})+\frac{1}{\pi}A(\lambda,c) \eeq 

with \(c\) a positive number with the dimension of \(\lambda\). That branch is chosen where the logarithm is real. Then the function \(A(\lambda,c)\) is a real and rational analytic function of \(\lambda\), finite around \(\lambda=0\) and which for negative \(\lambda\) is sufficiently positive. Now we consider two cases:

\begin{enumerate}
\item \(\eta(0)>0\). Then for small \(\lambda\): \beq \alpha-\lambda-\pi\bar{\xi}(\lambda)=\alpha-\lambda-\pi\bar{\sigma}(\lambda)\approx\eta(\lambda)\ln(-\frac{\lambda}{c}) \eeq 

The "jump across the cut" of the integrand is then: \beq F(\lambda)\rightarrow\frac{1}{\eta(\lambda)\ln(-\frac{\lambda}{c})}-\frac{1}{\eta(\lambda)[\ln(-\frac{\lambda}{c})+2\pi i]}\rightarrow 0 \hspace{1cm}(\lambda\rightarrow 0) \eeq 
\item \(\eta(0)=0\). We have: 
\beq 
\eqalign{
\alpha-\lambda-\pi\bar{\xi}(\lambda)&=\alpha-\lambda-\pi\bar{\sigma}(\lambda)\cr
&=\alpha-\lambda+\eta(\lambda)\ln(-\frac{\lambda}{c})-A(\lambda,c) 
}
\eeq 
and we find then for small \(\lambda\): 
\beq
\eqalign{
F(\lambda)&=\frac{1}{\alpha-\lambda+\eta(\lambda)\ln(-\frac{\lambda}{c})-A(\lambda,c)}\cr
&\qquad -\frac{1}{\alpha-\lambda+\eta(\lambda)[\ln(-\frac{\lambda}{c})+2\pi i]-A(\lambda,c)}\cr&\approx \frac{2\pi i\eta(\lambda)}{[\alpha-\lambda-A(\lambda,c)]^{2}}\rightarrow 0 \hspace{1cm}(\lambda\rightarrow 0)
}
\eeq 
\end{enumerate}

The conclusion is that the transition matrix element behaves non-exponentially but goes faster to zero than \(t^{-1}\) for large \(t\). Correspondingly the survival rate goes slower than exponential but faster than \(t^{-2}\) to zero for large \(t\).

Of course, the anomaly has been demonstrated only in a simplified model which however is realistic enough to warrant the expectation that the anomaly is characteristic for a much wider class of models. It may even be unavoidable.

\vspace{3cm}

{\bf Acknowledgements}

The author wishes to thank Prof. R. Kleiss, Dr. Th. Rijken and Chr. Dams for useful comments and suggestions.
\newpage

\appendix

\section{Appendix}

In this Appendix we prove the three theorems on Hilbert transforms mentioned in the text.

\begin{enumerate} 
\item Let \(\sigma(y)\) be the Hilbert transform of \(\eta(x)\): \beq \frac{1}{\pi}\int_{-\infty}^{\infty}\eta(x)P\frac{1}{x-y}dx=\sigma(y) \eeq then \(-\eta(y)\) is the Hilbert transform of \(\sigma(x)\): \beq \frac{1}{\pi}\int_{-\infty}^{\infty}\sigma(x)P\frac{1}{x-y}dx=-\eta(y) \eeq 

Proof: In the text we have already proven that \beq \frac{1}{\pi^{2}}\int_{-\infty}^{\infty}P\frac{1}{x-z}P\frac{1}{y-z}dz=\delta(x-y) \eeq 

We now have: 
\beq
\eqalign{
&\int_{-\infty}^{\infty}\eta(x)\delta(x-y)dx \cr
&= \frac{1}{\pi^{2}}\int_{-\infty}^{\infty}[\int_{-\infty}^{\infty}\eta(x)P\frac{1}{x-z}dx]P\frac{1}{y-z}dz=\eta(y) 
}
\eeq 

and therefore: 
\beq \eta(y)=\frac{1}{\pi}\int_{-\infty}^{\infty}\sigma(z)P\frac{1}{y-z}dz=-\frac{1}{\pi}\int_{-\infty}^{\infty}\sigma(z)P\frac{1}{z-y}dz \eeq 

\item The function: \beq \xi(z)\stackrel{\rm def}{=}\frac{1}{2\pi i}\int_{-\infty}^{\infty}\frac{\sigma(x)+i\eta(x)}{x-z}dx \eeq 

is zero in the lower half \(z\)-plane, analytic an regular in the upper half \(z\)-plane and has the property: \beq \lim_{z\downarrow x_{0}}\xi(z)=\sigma(x_{0})+i\eta(x_{0}) \eeq 

Here \(x\) and \(x_{0}\) are points on the real axis.

Proof: The function is obviously analytic and regular in the upper and lower half \(z\)-plane. Since the real axis acts as a closed branch cut the two branches are not analytically connected. We have: 
\beq 
\eqalign{
\lim_{z\downarrow x_{0}}\xi(z)&=\frac{1}{2\pi i}\lim_{z\downarrow x_{0}}\int_{-\infty}^{\infty}\frac{\sigma(x)+i\eta(x)}{x-z}dx\cr
&=\frac{1}{2\pi i}\int_{-\infty}^{\infty}\frac{\sigma(x)+i\eta(x)}{x-x_{0}-i\epsilon}dx\cr
&=\frac{1}{2\pi i}\int_{-\infty}^{\infty}[\sigma(x)+i\eta(x)].[P\frac{1}{x-x_{0}}+\pi i\delta(x-x_{0})]dx\cr
&=\frac{1}{2i}[-\eta(x_{0})+i\sigma(x_{0})]+\frac{1}{2}[\sigma(x_{0})+i\eta(x_{0})]\cr
&=\sigma(x_{0})+i\eta(x_{0})
}
\eeq 

This proves the first part. We also have: 
\beq 
\eqalign{
\lim_{z\uparrow x_{0}}\xi(z)&=\frac{1}{2\pi i}\lim_{z\uparrow x_{0}}\int_{-\infty}^{\infty}\frac{\sigma(x)+i\eta(x)}{x-z}dx\cr
&=\frac{1}{2\pi i}\int_{-\infty}^{\infty}\frac{\sigma(x)+i\eta(x)}{x-x_{0}+i\epsilon}dx\cr 
&= \frac{1}{2\pi i}\int_{-\infty}^{\infty}[\sigma(x)+i\eta(x)].[P\frac{1}{x-x_{0}}-\pi i\delta(x-x_{0})]dx \cr
&=\frac{1}{2i}[-\eta(x_{0})+i\sigma(x_{0})]-\frac{1}{2}[\sigma(x_{0})+i\eta(x_{0})]=0
}
\eeq 

This means that the regular function in the lower half plane is zero on the real axis and must therefore be zero throughout the lower half plane.
\item The function: \beq \tilde{\sigma} (x,y)=\frac{1}{\pi}\int_{-\infty}^{\infty}\eta(z)P\frac{1}{z-x}P\frac{1}{z-y}dz \eeq 

satisfies the property: \beq \tilde{\sigma} (x,y)=\frac{\sigma(x)-\sigma(y)}{x-y}+\pi\eta(x)\delta(x-y) \eeq 

Proof: 
\beq
\eqalign{
\frac{1}{\pi}\int_{-\infty}^{\infty}&z\eta(z)P\frac{1}{z-x}P\frac{1}{z-y}dz\cr
&{}=\frac{1}{\pi}\int_{-\infty}^{\infty}[(z-x)+x]\eta(z)P\frac{1}{z-x}P\frac{1}{z-y}dz \cr
&{}=\frac{1}{\pi}\int_{-\infty}^{\infty}\eta(z)P\frac{1}{z-y}dz+x\frac{1}{\pi}\int_{-\infty}^{\infty}\eta(z)P\frac{1}{z-x}P\frac{1}{z-y}dz \cr
&{}=\sigma(y)+x\tilde{\sigma} (x,y)=\sigma(x)+y\tilde{\sigma} (x,y)
}
\eeq 

It follows from this that: \beq (x-y)\tilde{\sigma} (x,y)=\sigma(x)-\sigma(y) \eeq 

and from this we obtain: \beq \tilde{\sigma} (x,y)=\frac{\sigma(x)-\sigma(y)}{x-y}+A(x)\delta(x-y) \eeq 

The \(A(x)\) can be evaluated: 
\beq
\eqalign{
A(x)&=\int_{-\infty}^{\infty}\tilde{\sigma} (x,y)dy-\int_{-\infty}^{\infty}\frac{\sigma(x)-\sigma(y)}{x-y}dy\cr
	&=\frac{1}{\pi}\int_{-\infty}^{\infty}\eta(z)P\frac{1}{z-x}[\int_{-\infty}^{\infty}P\frac{1}{z-y}dy]dz\cr
		&\qquad-\sigma(x)\int_{-\infty}^{\infty}P\frac{1}{x-y}dy+\int_{-\infty}^{\infty}\sigma(y)P\frac{1}{x-y}dy  \cr
	&=\pi\eta(x) \cr
}
\eeq
\end{enumerate}

\end{document}